\documentclass[a4paper]{jpconf}
\usepackage{graphicx}
\usepackage{lmodern}
\begin{document}
\title{TREX-DM: a low background Micromegas-based TPC for low-mass WIMP detection}

\author{F.J.~Iguaz, J.G.~Garza,
F.~Aznar\footnote{Centro Universitario de la Defensa, Universidad de Zaragoza, Zaragoza, Spain.},
J.F.~Castel, S.~Cebri\'an, T.~Dafni, J.A.~Garc\'ia, I.G.~Irastorza,
A.~Lagraba, G.~Luz\'on and  A.~Peir\'o}

\address{Laboratorio de F\'isica Nuclear y Astropart\'iculas, Universidad de Zaragoza, Spain.}

\ead{iguaz@unizar.es}

\begin{abstract}
Dark Matter experiments are recently focusing their detection techniques in low-mass WIMPs,
which requires the use of light elements and low energy threshold.
In this context, we describe the TREX-DM experiment,
a low background Micromegas-based TPC for low-mass WIMP detection.
Its main goal is the operation of an active detection mass $\sim$0.3~kg,
with an energy threshold below 0.4 keVee and fully built with previously selected radiopure materials.
This work describes the commissioning of the actual setup situated in a laboratory on surface
and the updates needed for a possible physics run at the Canfranc Underground Laboratory (LSC) in 2016.
A preliminary background model of TREX-DM is also presented, based on a Geant4 simulation,
the simulation of the detector's response and two discrimination methods:
a conservative muon/electron and one based on a neutron source.
Based on this background model, TREX-DM could be competitive in the search for low-mass WIMPs.
In particular it could be sensitive, e.g., to the low-mass WIMP interpretation of the DAMA/LIBRA
and other hints in a conservative scenario.
\end{abstract}

\section{Introduction}
The experimental effort of Dark Matter searches~\cite{Baudis:2012lb} has focused for many years
on the search for Weakly Interactive Mass Particles (WIMPs) of relatively large masses (of around 50 GeV or larger).
This idea was supported by theoretical considerations, that identified the WIMP with the neutralino of SUSY extensions
of the Standard Model and by the fact that the best WIMP detector techniques
available were already suited for this mass range. However, some recent experimental efforts have
focused on low-mass WIMPs (10-20 GeV or less), due to a number of hints that could be interpreted in these terms
and the well-known DAMA/LIBRA claim~\cite{Bernabei:2013rb},
incompatible with results from other experiments in conventional scenarios,
but that might be reconciled in models invoking low-mass WIMPs~\cite{Savage:2009cs}.
For this physics, specific experiments with an intrinsic low energy threshold detection technique
and preferably light target nuclei are needed.

\medskip
One of the possible detection options is a Time Projection Chamber (TPC), as it has an intrinsic low energy threshold
($\sim$100~eV) and has access to a rich topological information.
Moreover, recent advances in radiopure Micromegas readout planes~\cite{Cebrian:2011sc},
in the selection of radiopure materials for TPCs~\cite{Aznar::2013fa}
and in electronics~\cite{Baron:2008pb,Baron:2010pb, Anvar:2011sa}
are improving the low-background prospects and scalability of Micromegas-based TPCs.
As part of the T-REX project\footnote{T-REX webpage: http://gifna.unizar.es/trex/},
a prototype to assess the feasibility of a low-mass WIMP detector has been developed: TREX-DM.
Its main strategy is based on accumulating mass by increasing the gas pressure
and on reaching an energy threshold well below 0.4~keVee (as already observed in \cite{Aune:2014sa}).
It is now being commissioned at T-REX laboratory
and is foreseen to be installed at Canfranc Underground Laboratory (LSC) for a physics run in 2016.
This article is a short introduction to~\cite{Iguaz:2015fi}
and the last of a series of conference's proceedings~\cite{Iguaz:2015fia,Iguaz:2015fib,Garza:2015jgg}.
In Sec.~\ref{sec:Detector} the detector and the main results of the comissioning are described,
as well as the changes foreseen for the version to be installed at the LSC.
In Sec.~\ref{sec:BackModel},
a first background model of TREX-DM and the prospects to detect low-mass WIMPs are shown.
We conclude in Sec.~\ref{sec:Conc}.

\section{Detector description and comissioning}
\label{sec:Detector}
\begin{figure}[htb!]
\centering
\includegraphics[width=50mm]{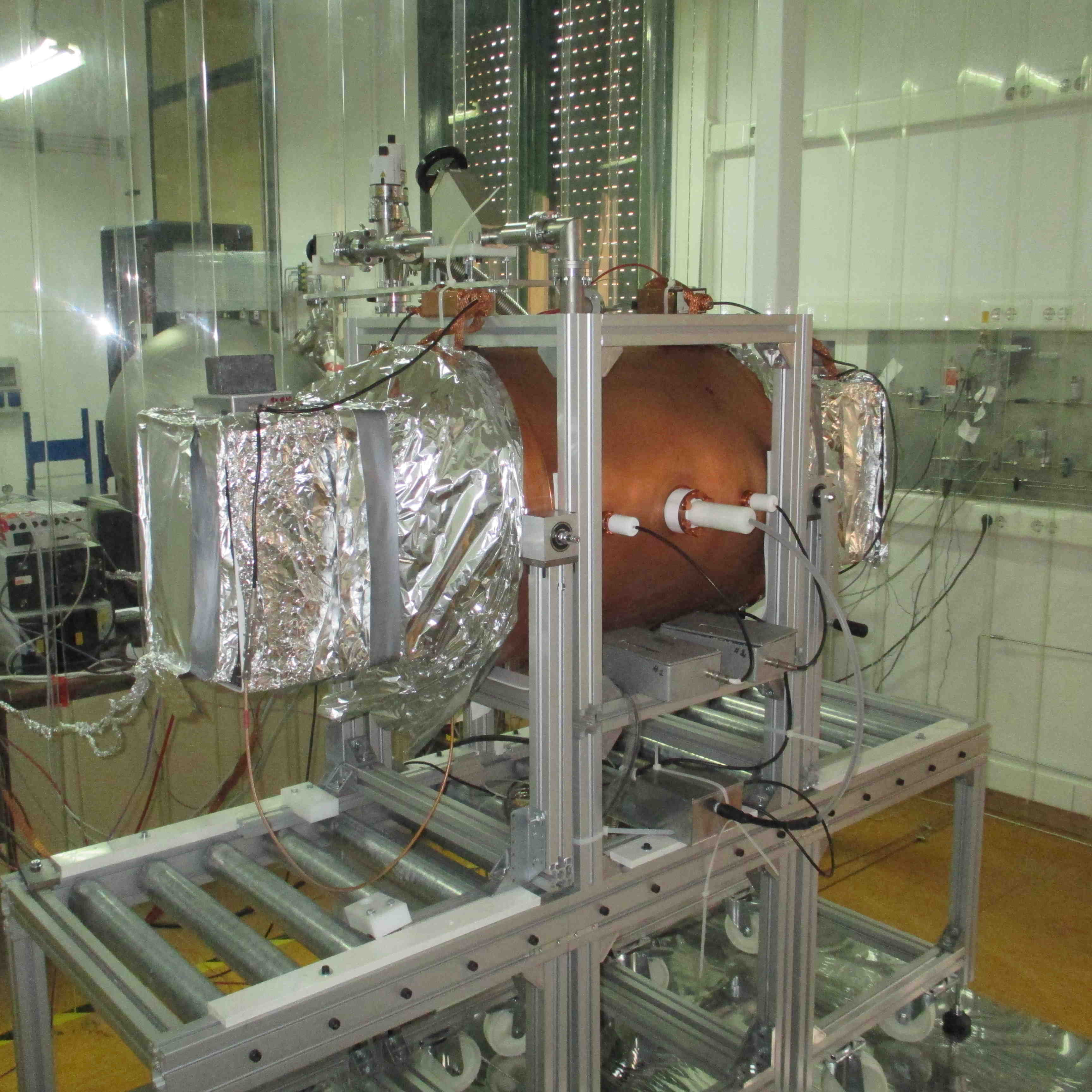}
\includegraphics[width=50mm]{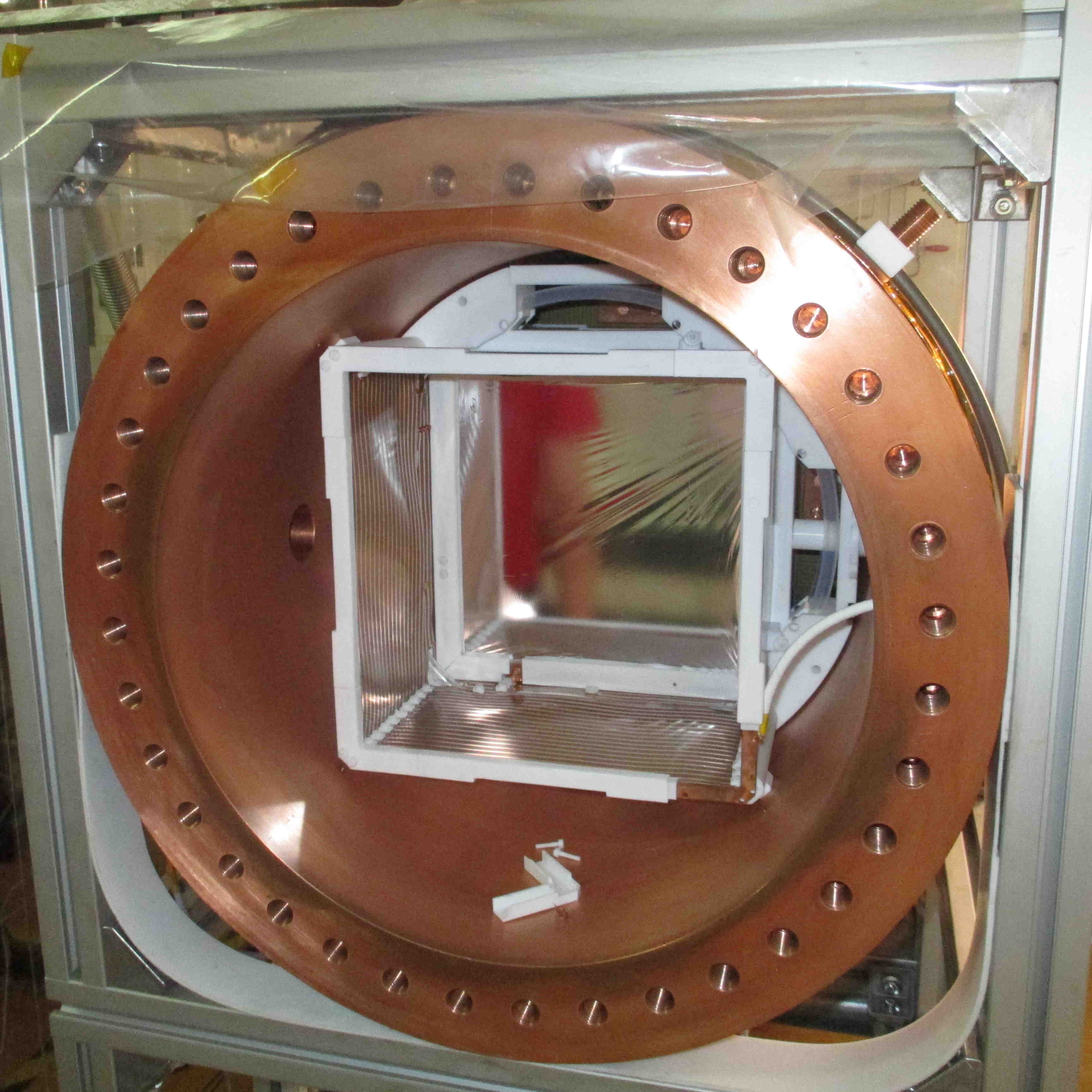}
\includegraphics[width=50mm]{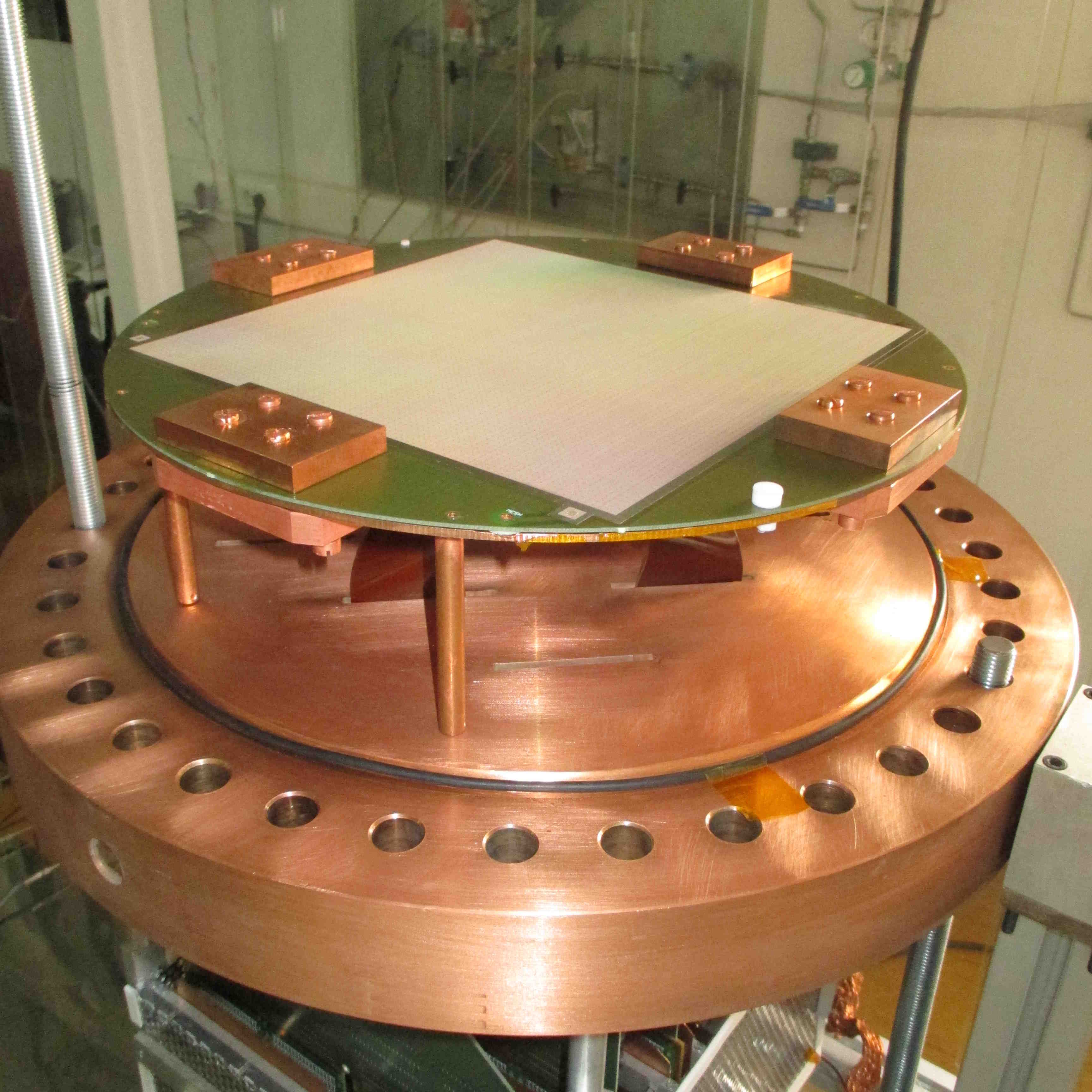}
\caption{Left: a view of the TREX-DM experiment during its commissioning.
Center: the inner part of the vessel, when one of the flat end caps has been removed. One can see the mylar foil
used as central cathode, the copper-kapton printed circuit used as field cage
and the copper squared ring, covered by a teflon gasket, where the drift chain ends.
Right: a Micromegas readout plane installed at its support base, which is screwed to one of the vessel's caps.
Four flat cables extract strip signals generated at the readout plane outside the vessel
by four feedthroughs and are then connected to four interface cards, and subsequently to four FEC cards.}
\label{fig:Setup}
\end{figure}

The TREX-DM detector (Fig.~\ref{fig:Setup}, left) has been designed to host 0.3~kg of argon mass at 10 bar
(or, alternatively, 0.16 kg of neon).
In some aspects, it is a scaled-up version of CAST Micromegas x-ray detectors~\cite{Aune:2014sa}
but with a $10^3$ larger active mass.
It is composed of a copper vessel, with an inner diameter of 0.5~m,
a length of 0.5~m and a wall thickness of 6~cm.
The vessel is divided in two active volumes by a central mylar cathode, which is connected to high voltage
by a tailor-made feedthrough.
At each side there is a 19~cm long field cage (Fig.~\ref{fig:Setup}, center),
composed of a copper-kapton printed circuit screwed to four teflon walls.
The copper strips are electrically linked one after the other by 10~M$\Omega$ resistors.
Each drift chain ends at a copper squared ring, connected to a customized high voltage feedthrough.
Its voltage is set by a variable resistor, in order to get a uniform field in the active volume.

\medskip
At each vessel cap, one Micromegas readout plane (Fig.~\ref{fig:Setup}, right) is screwed to a copper base,
which is then attached to the cap by four columns.
The readout has an active surface of $25.2 \times 25.2$ cm$^2$ and has been built by bulk technology~\cite{Giomataris:2006yg}.
Strip signals are extracted from the vessel by four flat cables.
Each flat cable comes out from the vessel through a slit at one of the vessel's caps
and a feedthrough screwed at the external side.
One of the cable's end is linked to one of the readout's footprints,
while the end situated outside the vessel is screwed to an interface card,
that routes the signals to an AFTER (ASIC For TPC Electronics Readout)-based
front-end card (FEC)~\cite{Baron:2008pb,Baron:2010pb}.

\medskip
Each active volume is calibrated by a $^{109}$Cd source (x-rays of 22.1 and 24.9~keV),
situated in an aluminum holder. This holder is screwed to a plastic rod that
enters into a teflon tube located inside the vessel through a leak-tight port.
The rod can be moved to four calibration points per active volume, situated at the corners of the central cathode.
Finally, there are two ports situated at the bottom and the top,
where gas enters and comes out from the vessel.
%The gas comes from a premixed bottle,
%whose pressure and flow is adjusted by a pressure transducer and a mass flowmeter via a slow control.

\medskip
The Micromegas readout planes have been characterized in Ar+2\%iC$_4$H$_{10}$ for pressures between 1.2 and 10~bar,
in steps of 1~bar with the help of a $^{109}$Cd.
Both detectors show a wide range of drift fields for which the mesh is transparent to primary electrons,
similar to those of other bulk detectors~\cite{Iguaz:2011fa}.
Both detectors show a similar gain for the same mesh voltage.
However, their maximum operation gain decreases with the gas pressure,
from $3\times10^3$ at 1.2~bar down to $5\times10^2$ at 10 bar.
There is also a degradation in terms of energy resolution, from 16\%~FWHM at 22.1~keV at 1.2~bar to 25\%~FWHM at 10~bar.
We cannot discard that this effect may be caused by the low quantity of quencher (2\%) in the gas,
as better results have been obtained in Ar+5\%iC$_4$H$_{10}$.
In the best noise conditions, the measured energy threshold was 1.0~keV at 1.2~bar and 1.4~keV at 3~bar.
These values do not reach the 0.4~keV obtained with other Micromegas detectors \cite{Aune:2014sa}.

\medskip
In the next months, the readout planes will be characterized in other gases:
Ar+5\%iC$_{4}$H$_{10}$, to study the role of isobutane at high pressure; and Ne+2\%iC$_{4}$H$_{10}$,
as higher gains and a better energy resolution have been measured in other Micromegas detectors~\cite{Iguaz:2012fi}.
In parallel, several changes are foreseen for the version to be installed at the LSC:
\begin{itemize}
 \item \textbf{Readout planes}: The actual readout planes are on a Printed Circuit Board (PCB).
 One of their core materials is FR4/phenolic, which is not clean in terms of radiopurity. To avoid this material,
 two new radiopure readouts are being designed,
 based on the latest Micromegas technologies: bulk~\cite{Giomataris:2006yg} and microbulk \cite{Andriamonje:2010sa}.
 \item \textbf{Electronics}: In the current DAQ implementation, the trigger is built out of the mesh signal.
 The energy threshold is thus limited by the electronic noise of the mesh channel,
 that is high due to its high capacitance ($\sim$5~nF).
 In the final DAQ, based on the AGET chip \cite{Anvar:2011sa},
 the trigger will be generated by strip signals,
 which could show a lower noise level due to their lower capacitance ($\sim$200~pF).
 \item \textbf{Shielding \& support}: The actual aluminum support structure will be replaced by a new copper-based one,
 as aluminum is tipically not radiopure. The structure should be compatible with a shielding,
 to reduce the effect of the external gamma and neutron flux.
 Other systems will also be affected like the gas and vacuum systems,
 whose components near the vessel should be rebuilt in copper to improve them in radiopurity terms.
\end{itemize}

\section{Background model of TREX-DM at LSC}
\label{sec:BackModel}
A first background model of the experiment, as it were installed and in operation at the LSC, has been created
to estimate the sensitivity of TREX-DM to low-mass WIMPs. We have considered two light gas mixtures at 10 bar:
Ar+2\%iC$_4$H$_{10}$ and Ne+2\%iC$_4$H$_{10}$, which are good candidates to detect low-mass WIMPs
and give a total active mass of 0.30 and 0.16~kg respectively.
This model is based on two pillars: a screening program of all materials
used in the detector and the simulation of the detector's response.
On one side, a wide range of materials and components used in TREX-DM detector has been measured
in terms of radiopurity: the vessel, the field cage, the shielding, the readout planes and the electronics acquisition system.
This screening program is mainly based on germanium gamma-ray spectrometry
performed at the LSC and, complementing these results,
on Glow Discharge Mass Spectrometry (GDMS) ~\cite{Cebrian:2011sc, Aznar::2013fa}.
On the other side, the simulation of the detector's response is based
on a Geant4 program~\cite{Agostinelli:2003sa}
and REST code~\cite{Tomas:2013at}. This last code simulates all the specific TPC processes:
from the creation of primary electrons until the generation of signals at mesh and strips.

\begin{figure}[htb!]
\centering
\includegraphics[width=80mm]{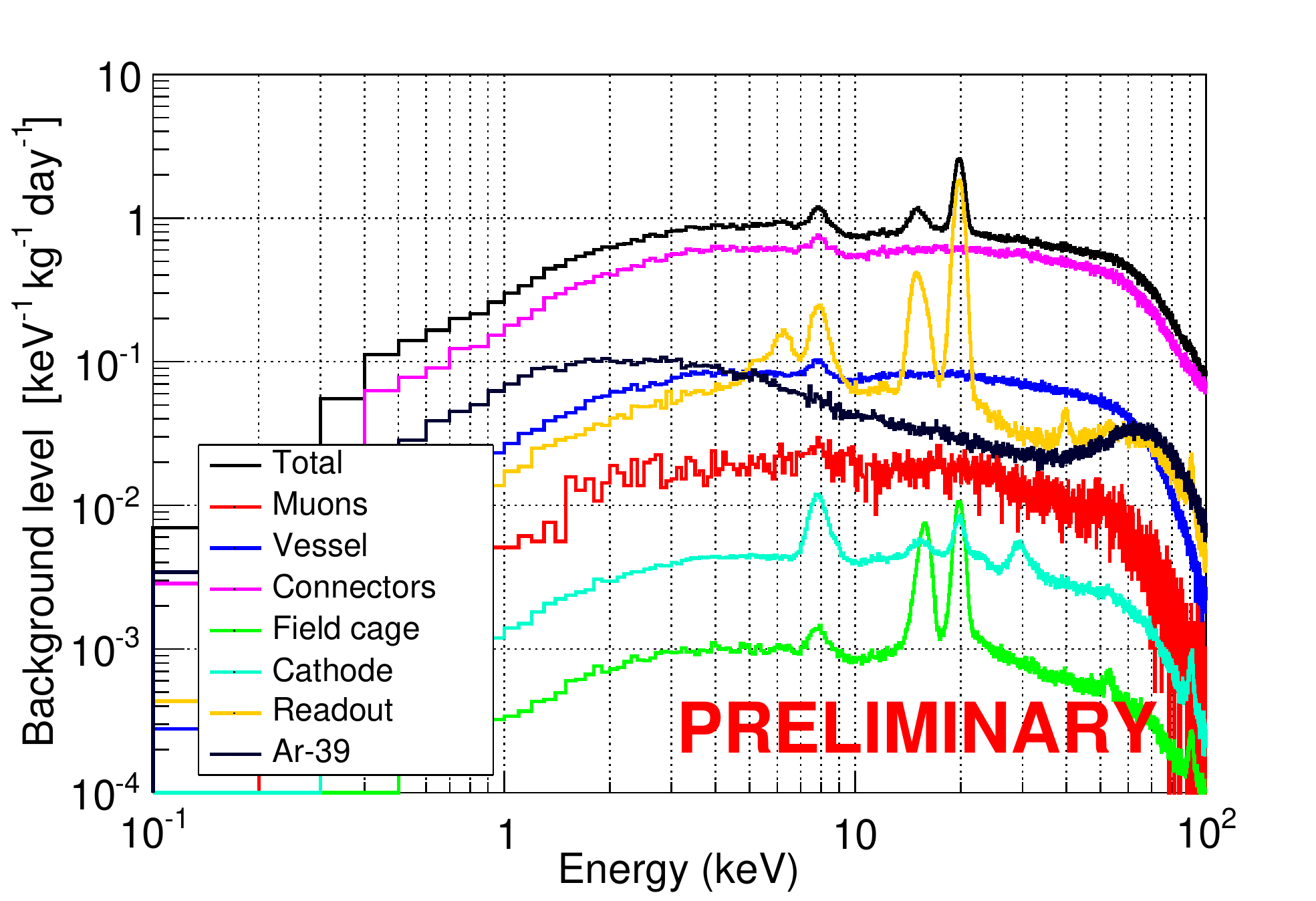}
\includegraphics[width=62mm]{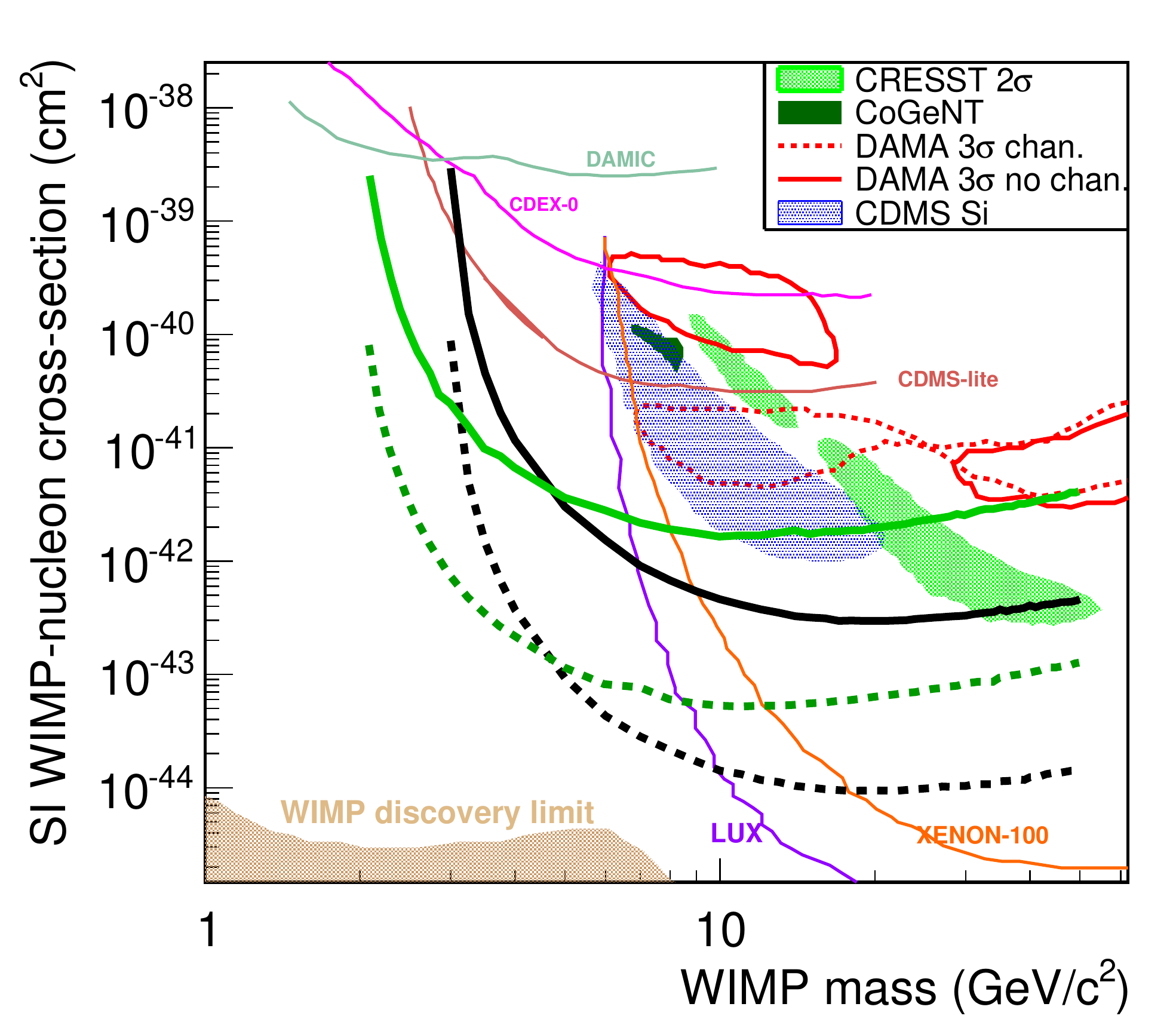}
\caption{Left: Background spectrum expected in TREX-DM (black line) during a physics run
if operated in Ar+2\%iC$_4$H$_{10}$ at 10 bar.
The contribution of the different simulated components is also plotted: external muon flux (red line),
vessel contamination (blue line), connectors (magenta line), field cage (green line), central cathode (brown line),
Micromegas readout planes (orange line) and $^{39}$Ar (dark blue line).
Right: WIMP parameter space focused on the low-mass range. Filled regions represent the values that may explain
the hints of positive signals observed in CoGeNT, CDMS-Si, CRESST and DAMA/LIBRA experiments. The thick lines are
the preliminary sensitivity of TREX-DM supposing a 0.4~keVee energy threshold and two different hypothesis
on background and exposure: 100 (solid) and 1(dashed) keV$^{-1}$ kg$^{-1}$ day$^{-1}$,
and 1 and 10 kg-year respectively, and for both argon- (black) and neon-based mixtures (green).
Further details can be found in~\cite{Iguaz:2015fi}.}
\label{fig:BackSpec}
\end{figure}

\medskip
The radioactive isotopes of the most relevant prototype's components have been simulated,
and the results have been scaled by the activities of the screening program.
In some cases, a radiopure version has been considered, as in the case of the readout planes,
or some references in literature have been used, if they are more sensitive than germanium gamma-ray spectrometry.
Finally, for the specific case of argon-based mixtures and the $^{39}$Ar isotope,
the activity in~\cite{Agnes:2015pa} has been used. Two analysis have been considered: the first one is a modified
version of CAST analysis \cite{Aune:2014sa},
optimized to discriminate low energy x-rays from complex topologies like gammas and cosmic muons.
It is particularly useful for this detector as WIMP-induced nuclear recoils and low energy x-rays
will show similar features at high gas pressure. The second one is based on the simulation of a $^{252}$Cf neutron source,
which reproduces better the expected WIMP signals and has been used to discard any artificial fiducial cut.
Fixing a signal efficiency of 80\%, the expected background level is $\sim$1 keV$^{-1}$ kg$^{-1}$ day$^{-1}$
in 2-7~keV for both argon and neon-based mixtures, as shown in Fig.~\ref{fig:BackSpec} (left).
The main contribution is due to the connectors (70-80\%),
followed by those of the readout planes, the vessel and the $^{39}$Ar isotope in the case of argon.
Supposing a 0.4~keVee energy threshold and this background level, TREX-DM could be sensitive with an exposure of 1 kg-year
to a relevant fraction of the low-mass WIMP parameter space (Fig.~\ref{fig:BackSpec}, right),
including the regions invoked in some interpretations of the DAMA/LIBRA hint.

\medskip
From this first background model, we have also obtained some conclusions for the setup to be installed at the LSCs:
\begin{itemize}
 \item \textbf{Gas}: in the case of argon, it should be extracted from underground sources
 because the atmospheric one may increase the $^{39}$Ar contribution a factor $10^3$.
 \item \textbf{Connectors}: they should be shielded with a 1 cm-thick layer of copper and a 1 cm-thick layer of lead.
 To reach lower values, they should be put further away for active volumes.
 \item \textbf{Readout planes}: they should be replaced by a radiopure version of the bulk technology
 or by the microbulk one. Note that the activity of the actual readout planes is a factor $10^4$ higher than
 the simulated one.
 \item \textbf{Outer shielding}: it should be designed to reduce the contribution of the external gamma flux
 and neutrons (including muon-induced in the shielding or in the rock) to this level.
\end{itemize}

\section{Conclusions}
\label{sec:Conc}
TREX-DM is a low background Micromegas-based TPC for low-mass WIMP detection. Its main goal is the operation of
a light gas at high pressure (either $\sim$0.3~kg of argon or $\sim$0.16~kg of neon)
with an energy threshold of 0.4~keVee or below
and fully built with previously selected radiopure materials. The detector is being commissioned at the T-REX laboratory
and several changes are planned for a physics run at LSC during 2016.

%%%%%%%%%%%%%%%%%%%%%%%%%%%%%%%%%%%%%%%%%%%%%%%%%%%%%%%%%%%%%%%%%%
\ack
We acknowledge the Micromegas workshop of IRFU/SEDI for bulking our readout planes and the
\emph{Servicio General de Apoyo a la Investigaci\'on-SAI} of the University of Zaragoza for the fabrication
of many mechanical components.
We also thank D.~Calvet from IRFU/SEDI for his help with the AFTER electronics.
We acknowledge the support from the European Commission under the European Research Council
T-REX Starting Grant ref. ERC-2009-StG-240054 of the IDEAS program of the 7th EU Framework Program,
the Spanish Ministry of Economy and Competitiveness (MINECO) under grants FPA2011-24058 and FPA2013-41085-P.
%and the University of Zaragoza under grant JIUZ-2014-CIE-02.
F.I. acknowledges the support from the \emph{Juan de la Cierva} program
and T.D. from the \emph{Ram\'on y Cajal} program of MINECO.

%%%%%%%%%%%%%%%%%%%%%%%%%%%%%%%%%%%%%%%%%%%%%%%%%%%%%%%%%%%%%%%%%%


\section*{References}
\begin{thebibliography}{99}
%%%%%%%%%%%%%%%%%%%%%%%%%%%
%%% Introduction        %%%
%%%%%%%%%%%%%%%%%%%%%%%%%%%
%%%% Reviews of Dark Matter.
\bibitem{Baudis:2012lb}
Baudis L
2012
%\emph{Direct dark matter detection: The next decade}
{\it Physics of the Dark Universe} {\bf 1} 94-108.
%%%% DAMA signal.
\bibitem{Bernabei:2013rb}
Bernabei R \emph{et al.}
2013
%\emph{Final model independent result of DAMA/LIBRA-phase 1}
\textit{Eur. Phys. J. C} \textbf{73} 2648.
%%%% Compatibility of DAMA/LIBRA with other results.
\bibitem{Savage:2009cs}
Savage C
2009
%\emph{Compatibility of DAMA/LIBRA dark matter detection with other searches}
\textit{JCAP} \textbf{0904} 010.
%% Radiopurity of Micromegas readouts.
\bibitem{Cebrian:2011sc}
Cebri\'an S \emph{et al.}
2011
%\emph{Radiopurity of micromegas readout planes},
{\it Astropart. Phys.} {\bf 34} 354.
%% Radiopurity of other components.
\bibitem{Aznar::2013fa}
Aznar F \emph{et al.}
2013
%\emph{Assesment of material radiopurity for Rare Event experiments using Micromegas},
{\it JINST} {\bf 8} C11012.
%% AFTER elecrtronics.
\bibitem{Baron:2008pb}
Baron P \emph{et al.}
2008
%\emph{AFTER, an ASIC for the Readout of the Large T2K Time Projection Chambers},
\textit{IEEE Trans. Nucl. Sci.} \textbf{55} 1744.
%% Second reference.
\bibitem{Baron:2010pb}
Baron P \emph{et al.}
2010
%\emph{Architecture and implementation of the Front-End Electronics of the Time Projection Chambers in the T2K Experiment},
\textit{IEEE Trans. Nucl. Sci.} \textbf{57} 406.
%% AGET electronics.
\bibitem{Anvar:2011sa}
Anvar S \emph{et al.}
2011
%\emph{AGET, the GET front-end ASIC, for the readout of the time projection chambers used in nuclear physics experiments},
\textit{IEEE NSS/MIC} 745.
%% CAST detectors.
\bibitem{Aune:2014sa}
Aune S \emph{et al.}
2014
\textit{JINST} \textbf{9} P01001.
%%%%%%%%%%%%%%%%%%%%%%%%%%%
%%% TREX-DM papers      %%%
%%%%%%%%%%%%%%%%%%%%%%%%%%%
%% Main article.
\bibitem{Iguaz:2015fi}
Iguaz F J \emph{et al.}
2015
%\emph{TREX-DM: a low background Micromegas-based TPC for low-mass WIMP detection}
\textit{submitted to Eur. Phys. J. C} \textit{Preprint} physics.ins-det/1512.01455.
%% LTPC proceeding.
\bibitem{Iguaz:2015fia}
Iguaz F J \emph{et al.}
2015
%\emph{proceeding of the 7th Symposium on Large TPCs for Low-Energy Rare Event Detection}
%\textit{Preprint} physics.ins-det/1503.07085.
\textit{J. Phys. Conf. Ser.} \textbf{650} 012005.
%% Patras-2015.
\bibitem{Iguaz:2015fib}
Iguaz F J \emph{et al.}
2015
\emph{Proc. PATRAS 2015} (Hamburg: Verlag Deutsches Elektronen-Synchrotron) p 100
\textit{Preprint} physics.ins-det/1509.02035.
%% MPGD-2015.
\bibitem{Garza:2015jgg}
Garza J G {\it et al.}
2015
\emph{talk at the MPGD 2015 conference}, Trieste (Italy).
%\emph{Proceedings of the MPGD 2015 conference}
%\textit{Preprint} physics.ins-det/1511.XXXX.
%%%%%%%%%%%%%%%%%%%%%%%%%%%
%%% Description         %%%
%%%%%%%%%%%%%%%%%%%%%%%%%%%
%% Bulk technology.
\bibitem{Giomataris:2006yg}
Giomataris I \emph{et al.}
2006
%\emph{Micromegas in a bulk},
\textit{Nucl. Instrum. Meth. A} \textbf{560} 405.
%% MIMAC detector.
\bibitem{Iguaz:2011fa}
Iguaz F J \emph{et al.}
2011
%\emph{Micromegas detector develpments for Dark Matter directional detection with MIMAC},
\textit{JINST} \textbf{6} P07002.
%% Low quantity of quencher.
\bibitem{Iguaz:2012fi}
Iguaz F J, Ferrer-Ribas E, Giganon A and Giomataris I
2013
\textit{JINST} \textbf{7} P04007.
%% Microbulk technology
\bibitem{Andriamonje:2010sa}
Andriamonje S \emph{et al.}
2010
%\emph{Development and performance of Microbulk Micromegas detectors}, 
\textit{JINST} \textbf{5} P02001.
%%%%%%%%%%%%%%%%%%%%%%%%%%%%
%%% Background model     %%%
%%%%%%%%%%%%%%%%%%%%%%%%%%%%
%%% Reference to Geant4.
\bibitem{Agostinelli:2003sa}
Agostinelli S \emph{et al.}
2003
\textit{Nucl. Instr. Meth. A} \textbf{506} 250.
%%% Reference to REST.
\bibitem{Tomas:2013at}
Tomas A
2013
\textit{JINST} \textbf{TH} 001, CERN-THESIS-2013-062.
%%% DarkSIDE, last paper.
\bibitem{Agnes:2015pa}
Agnes P \emph{et al.}
2015
\textit{Preprint} astro-ph/1510.00702.
\end{thebibliography}
\end{document}